# VISCOSITY AND RELAXATION PROCESSES OF THE LIQUID BECOME AMORPHOUS Al-Ni-REM ALLOYS

V.I. Lad`yanov, A.L. Bel`tyukov, S.G. Menshikova[1], M.G. Vasin

Physical-Technical Institute, Ural Branch of Russian Academy of Sciences, 426000,

132 Kirov Str., Izhevsk

**Abstract:** The temperature and time dependencies of viscosity of the liquid alloys, $Al_{87}Ni_8Y_5$, $Al_{86}Ni_8La_6$, $Al_{86}Ni_8Ce_6$, and the binary Al-Ni and Al-Y melts with Al concentration over 90 at.% have been studied. Non-monotonic relaxation processes caused by destruction of non-equilibrium state inherited from the basic-heterogeneous alloy have been found to take place in Al-Y, Al-Ni-REM melts after the phase solid-liquid transition. The mechanism of non-monotonic relaxation in non-equilibrium melts has been suggested.

*Keywords*: melt, viscosity, non-equilibrium and equilibrium state of the melt, relaxation.

## 1. Introduction

Al-based (80-90 at.%) alloys containing transition and rare-earth metals (TM and REM) are highly disposed towards amorphization. Such alloys in the amorphous state have high strength, ductility and corrosion resistance in comparison with the properties of cast industrial alloys [1,2]. The possibility to influence the alloys structure formation at ultra quenching, i.e., in the melt obtaining stage, by temperature-time treatment allows further amorphization ability and other properties improvement [3,4]. On studying structure and physical-chemical properties of liquid multi-component alloys, Al-TM-REM included, numerous experimental results have been obtained. They testify to the fact that on heating the alloy structures are greatly changed (e.g. [5]) and after melting the process of relaxation in melts take a long time [4,5]. Temperature and time

---

[1]Corresponding authors at: Physical-Technical Institute, Ural Branch of Russian Academy of Sciences, 132 Kirov Str. Izhevsk, *E-mail address*: svetlmensh@mail.ru (S. Menshikova)

dependencies of the liquid melts properties, e.g. viscosity (ν) as the most structural sensitive one, are often used as a structural transformation indicator.

Earlier in the papers [9-12] in the study of the time dependencies of viscosity of liquid alloys of the Al-Ni-(Ce/La/Y) and more complex compositions Al-(Ni,Co)-(Gd,Y/Tb) the long non-monotonic relaxation processes in the melts after the phase solid-liquid transition have been found. The nature of long-term relaxation of the melts and causes of non-monotonic of this process remain unclear. Further study of this issue is necessary.

The experimental results on the temperature and time dependencies of kinematic viscosity in the liquid $Al_{87}Ni_8Y_5$, $Al_{86}Ni_8(La/Ce)_6$ alloys as well as the melts of the binary Al-Ni and Al-Y with Al concentration over 90 at.% have been presented in this paper.

## 2. Objects and experimental procedure

Al-Ni and Al-Y alloys were produced by alloying Al (A999) and $Al_{85}Ni_{15}$ or $Al_{90}Y_{10}$ in a viscosimeter furnace in purified helium at $1100^oC$ being held isothermally at least for 1 hour. $Al_{85}Ni_{15}$ was obtained by alloying Al (A999) and electrolytic Ni (99.6 mass.%) in a resistance furnace at the residual pressure of $10^{-2}$ Pa, at $1670^oC$ for 30 minutes. $Al_{90}Y_{10}$, $Al_{87}Ni_8Y_5$ and $Al_{86}Ni_8(La/Ce)_6$ were obtained in an electric arc furnace under purified argon conditions[2]. The initial components for melting these alloys were the elements with the content of the basic metal: aluminium 99.99, lanthanum 99.99, yttrium 99.8, cerium 99.9 and electrolytic nickel 99.6 mass.%.

Viscosity measurements were carried out by the damped torsional vibrations method. To calculate viscosity the logarithmic decrement and the period of vibrations of the suspended system of viscometer with the crucible filled with the test sample were measured. The experimental device and the method of calculation of viscosity are described in detail in [13,14]. Cylindrical cups made of $Al_2O_3$ were used as crucibles. To exclude an uncontrolled influence of the oxide film developed on the alloy surface upon the measurements results a lid was placed on the sample top used as the second face surface [15]. The lids were produced from $Al_2O_3$ cups, the diameter being 0.5 mm less than the inner crucible diameter. During experiment the lid fits tightly to the top border of the melt and moves together with the crucible in the process of torsional vibrations. The crucibles and the lids were pre-annealed in the vacuum furnace at 1650°C for 1 hour, the residual pressure being equal to $10^{-2}$ Pa. Before performing the measurements each sample was remelted at 1200°C in a crucible covered with a lid in a viscosimeter furnace and then cooled to room temperature. All measurements were performed in a protective atmosphere of purified helium.

---

[2] $Al_{90}Y_{10}$, $Al_{87}Ni_8Y_5$ and $Al_{86}Ni_8(La/Ce)_6$ were produced at the G.V. Kurdyumov Institute of Metallophysics, NAS, Ukraine.

Alloys liquidus temperatures were determined by differential thermal analysis (DTA) and controlled by the temperature dependence of the logarithmic decrement of the suspended system of viscometer with the test sample. The decrement values when heating the alloy from the solidus temperature to the liquidus temperature generally increase more than 10 times. A further increase in temperature (above the liquidus) for "low viscous" fluids, which include metallic melts [16], leads to a monotonic decrease in the values of the decrement. The thermo-gram DTA of the $Al_{87}Ni_8Y_5$ alloy and the temperature dependence of the decrement, obtained in the heating regime, are shown in Fig. 1. The liquidus temperatures of the alloys are shown in Table.

At viscosity measurements the general mean error was no greater than 4% and the unit error was no more than 2.5%.

**3. Results and discussion**

Fig. 2 presents the temperature dependence of viscosity of the $Al_{87}Ni_8Y_5$ melt, obtained on heating (line 1) and cooling (2) in the range from the liquidus temperature to 1200°C, the steps being equal to 30°C and 50°C, accordingly. The isothermal exposures at each temperature were as long as 30 minutes. Under these conditions on heating a sharp decrease of the viscosity values in the vicinity of 1040°C and on cooling below 1040°C hysteresis was observed in the polytherm. The hysteresis is negative, i.e. the viscosity values, obtained in the cooling regime, are lower the values obtained in the heating regime.

Fig. 3 shows the time dependencies of viscosity $\nu(\tau)$ of the $Al_{87}Ni_8Y_5$ melt at different temperatures. These dependencies were obtained under isothermal conditions (the temperature changes was no more than $\pm 5^oC$) after heating the sample from room up to the given temperature at the speed of 30-40 degrees/minute. It is seen in Fig. 3 that the equilibrium reaching process in the melt was of a non-monotonic character. The $\nu(\tau)$ dependencies are presented in the complicated forms: as the viscosity values decrease with time a clearly defined peak appears in the isotherm. As the melt overheating temperature compared to the liquidus line goes up the relative value of the peak as well as the value of the time at which the peak appears go down. Similar time dependencies of viscosity were obtained earlier for liquid ternary alloys of Al-Ni-(Ce/La), and more complex compositions of Al-(Ni,Co)-(Gd,Y/Tb) [9-12]. As an example, Fig. 4 shows the data of the $Al_{86}Ni_8La_6$ and $Al_{86}Ni_8Ce_6$ melts at 1100°C. Taking the time dependencies of viscosity into consideration the values of the melts relaxation time as the time of reaching the constant values of viscosity of the melts were determined. The time of liquid alloys relaxation in the vicinity of the liquidus temperature was about 200 minutes. With the increase of the melt temperature the time of melt relaxation decreases (see Table).

Thereafter, the temperature dependencies of the melt viscosity were measured on heating and cooling after a long isothermal exposure at the initial temperature. Before taking measurements the sample was heated to the given initial temperature at 30-40 degrees/minute and then it was held for a long time. The exposure time exceeded the melt relaxation time at the given temperatures. Further viscosity measurements were taken after the sample had been isothermally held at each temperature for 15 minutes. In these experimental conditions the temperature dependencies of viscosity of the $Al_{87}Ni_8Y_5$ melt on heating (after holding at 870° C for 300 min) and cooling (after holding at 1100° C for 90 min) were obtained. Besides, the temperature dependencies of viscosity of the $Al_{86}Ni_8La_6$ and $Al_{86}Ni_8Ce_6$ melt on cooling after long isothermal holding at 900°C, 1000°C, 1100°C (for alloy $Al_{86}Ni_8La_6$) and 1050°C, 1200°C (for alloy $Al_{86}Ni_8Ce_6$) were obtained. The temperature dependencies of viscosity of the $Al_{87}Ni_8Y_5$ obtained (polytherms 3 and 4, Fig. 2) practically coincided. They were of a monotonic character and correlated well with the polytherm obtained on cooling (Fig. 2, curve 2). The temperature dependencies of viscosity for the $Al_{86}Ni_8La_6$ and $Al_{86}Ni_8Ce_6$ liquid alloys were also in good agreement (Fig. 5). All of the temperature dependencies of the viscosity, obtained in these experimental conditions, are monotonic and are satisfactorily described by the Arrhenius equation:

$$v = A \cdot \exp\left(\frac{E_v}{RT}\right) \qquad (1)$$

where A is a coefficient; $E_v$ is the energy of viscous flow activation; R is the molar gas constant and T is the absolute temperature. The parameters of the approximating equations of the viscosity polytherm are in the Table. Taking those data into consideration the relaxation processes in the melts are believed to occur after the phase solid-liquid transition.

To clarify the nature of the features observed in the polytherms and time dependencies of viscosity of the liquid $Al_{87}Ni_8Y_5$ alloy the melts of Al-Ni and Al-Y binary systems with an aluminium content of more than 90 at. % were studied.

The viscosity polytherms of the Al-Ni melts with nickel content of up to 10 at.% were presented in [17]. The temperature and time dependencies of viscosity of the $Al_{99}Y_1$ and $Al_{95}Y_5$ liquid alloys were presented in [18]. In addition, we measured the time dependencies of viscosity of the $Al_{95.5}Ni_{4.5}$ melt, as well as the temperature and time dependencies of viscosity of the liquid $Al_{90}Y_{10}$ alloy. To obtain the polytherms of viscosity the sample was heated and then cooled and isothermally held at each temperature for 20 minutes. The time dependencies were obtained by overheating the melt over the liquidus line, first it being heated from room temperature at 30-40 degree/min.

The temperature and time dependencies of viscosity of the Al-Ni melts are given in Figs. 6

and 7. The Al-Ni viscosity polytherms obtained at heating and cooling coincide (no hysteresis was observed). They are of a monotonic character (Fig. 6). As seen from Fig. 7 the melt viscosity values at constant temperature does not depend on the time of exposure. Prolonged relaxation processes in the melts of this system have not been observed.

It is shown in (Fig. 8) that at heating the values of viscosity decrease rather sharply in the vicinity of the liquidus line and on further cooling hysteresis is negative in the Al-Y temperature viscosity dependencies. For the melts with the 1 at.% ittrium concentration viscosity hysteresis starts on cooling the sample in the area of 30-50°C higher than the liquidus line. If the Y concentation is 5 and 10 at.% hysteresis is observed to start at 1100°C.

The time dependencies of viscosity of the Al-Y melt are given in Fig. 9. At 700°C the $Al_{99}Y_1$ liquid melt viscosity decreases non-monotonically with time (Fig. 9, line 1). The stable values at this temperature are found to observe in 100-120 minutes from the beginning of the isothermal exposure. No change of viscosity is discovered in $Al_{99}Y_1$ isothermally held at 800°C.

Prolonged abnormal growth of the calculated values of ν of the $Al_{95}Y_5$ and $Al_{90}Y_{10}$ melts at the beginning of isothermal exposures (lines 2, 4 and 5 in Fig. 10), in our opinion, is not associated with an increase in viscosity, but is a consequence of the growth of the values of decrement of suspension system viscometer with test sample. Apparently, at the beginning of isothermal exposure these alloys are hetero-phase systems. Each of these systems consist of a liquid and the crystals of $Al_3Y$ intermetallide. In this case, as noted above, growth of the decrement values due to changes in the ratio of liquid and solid phases in the investigated alloy at its melting. Used in the method for measuring the viscosity does not apply in the study of hetero-phase systems [16], but this form of presentation of the results clearly reflect the processes occurring in the melt. After the end of melting there is a long time of non-monotonic reduction in the values of ν in the time dependencies of viscosity of the $Al_{95}Y_5$ melt at 870° C (Fig. 9, line 2) and $Al_{90}Y_{10}$ melt at temperatures of 1000 °C (Fig. 9, 3) and 1200 °C (Fig. 9, 4). Non-monotonic is manifested most clearly in the temperature dependencies of viscosity of the Al$_{90}$Y$_{10}$ melt. Those time dependencies are similar qualitatively on the time dependencies of viscosity of the liquid Al-TM-REM alloys discussed earlier (Figs. 3 and 4). It should also be noted that the viscosity values of the Al-Y melts established during the isothermal exposure are in good agreement with the values of ν in polytherms obtained in at cooling (Fig. 8).

The relaxation processes and hysteresis of the viscosity polytherms of both Al-Y and Al-TM-REM are likely to be caused by the irrevesible decomposition of the non-equilibrium state of the melts inherited from the basic-heterogeneous alloy [12]. It is supposed that the non-equilibrium state in the liquid alloys is realized due to retention of the atomic micro-groups with the ordering type chemical compounds with high melting points. These compounds present in the initial solid alloys. It has been proven by the X-ray phase analysis that the initial alloys, Al$_{87}$Ni$_8$Y$_5$, Al$_{86}$Ni$_8$La$_6$, Al$_{86}$Ni$_8$Ce$_6$ consisted of three phases, α-Al (solid solution on aluminum-based), Al$_3$Ni and Al$_x$REM$_y$ compound (Al$_3$Y, Al$_{11}$La$_3$, Al$_{11}$Ce$_3$, respectively). In the investigated concentration range the initial binary Al-Ni alloys comprise two phases: α-Al and Al$_3$Ni. Also the initial binary Al-Y alloys comprise two phases: α-Al and Al$_3$Y.

To explain the results obtained it may be supposed that instantaneously on heating the Al-Y and Al-Ni-Y melts the areas which contain a relatively high-melting compound, Al$_3$Y, are broken down rather slowly. For them to be completely dissociated the Y concentration should be equalized by diffusion over the whole volume. Until it has occurred non-equilibrium atomic micro-groups (dispersed particles) may be formed by the non-dissolved complexes, Al$_3$Y. In this case the non-monotonic nature of the time dependencies in the melt viscosity is accounted for by the competitive influence of the two simultaneous decomposition processes, i.e. large micro-groups dispersion and the smallest micro-groups dissociation [12]. The former process (dispersion) results in the increase of the non-equilibrium concentration of the particles included in these micro-groups, the latter (dissociation) makes them decrease in quantity. This change can be presented in the form of a semi-empirical time dependence:

$$c(t,T) = \exp\left(-\frac{t}{\tau_{cut}}\right) \cdot \left[c^* - (c^* - c_0) \cdot \exp\left(-\frac{t}{\tau_{dis}}\right)\right], \qquad (2)$$

where $\tau_{dis} = \tau_0 \exp(E_{dis}/kT)$ is the dispersion time; $\tau_{cut} = \tau_0 \exp(E_{cut}/kT)$ is the dissociation time; and $c^*$ и $c_0$ are the maximum and the initial concentrations of the dispersed particles; $E_{dis}$ and $E_{cut}$ are the activation energies of dispersion and dissociation, accordingly; $k$ is Boltzmann constant.

According to Simha-Einstain equation [20,21] the melt viscosity, $\eta$, is influenced by the volume concentration of the dispersed particles, $\varphi$:

$$\eta \approx \eta_0(1+\lambda\varphi), \qquad (3)$$

where $\lambda$ is the form factor of the dispersed particles ($\lambda = 2.5$ in case of spherical particles). As long as dispersed particles fail to form a crystal structure, they exist in the form of clusters, their dimension, $x$, being different from the space dimension ($x < 3$), i.e. it is referred to as a fractal dimension. With their linear size, $L$, increase the micro-groups mass increases more slowly than $L^3$, $c \sim L^x$. This leads to the effective increasing of the disperse fraction volume, because the liquid around these clusters effectively associates with them. Thus

$$\varphi = c^{x/3}, \qquad (4)$$

and $\lambda > 2.5$. At the beginning of the isothermal melt exposure the dispersed particle concentration is fairly low and does not influence significantly on the viscosity. On dispersion when the concentration of the non-equilibrium particles reaches some maximum value, their influence becomes greater. With further Y and Al dissolution and the dispersed particles decrease in number their influence on the viscosity becomes lower again. Using (3), (4) and (2), the time dependence of viscosity may be represented as follows:

$$\eta(t) = \eta_0 \left[ 1 + \lambda \cdot e^{\frac{-tx}{3\tau_{cut}}} \cdot \left[ c^* - (c^* - c_0) \cdot e^{\frac{-t}{\tau_{dis}}} \right]^{\frac{x}{3}} \right] \quad (5)$$

The time dependencies of viscosity in the non-equilibrium melt, obtained according to (5) at various temperatures are given in Fig. 10. As seen in Fig. 3 the curves obtained are qualitatively well agreed with the time dependencies of viscosity obtained by experiment. In particular, as the melt temperature goes up the temperature dependence maximum is shifted to the area of lesser time dependencies, the height and width of the peak being reduced.

## 4. Conclusions

Non-monotonic relaxation processes in the melts, Al-Y and Al-Ni-REM, caused by decomposition of the initial non-equilibrium state have been found by the experimental investigations of the temperature and time dependencies of viscosity of the melts, $Al_{87}Ni_8Y_5$, $Al_{86}Ni_8La_6$, $Al_{86}Ni_8Ce_6$ and liquid binary alloys, Al-Ni and Al-Y. This state in the liquid melts was formed by inheritance of the atomic micro-groups with the $Al_xREM_y$ - type ordering on melting. In particular, the non-equilibrium state in the liquid Al-Y and $Al_{87}Ni_8Y_5$ melts was formed by inheritance of the atomic micro-groups with ordering type of $Al_3Y$.

For the melts to be at the equilibrium state they should be isothermally held for a long period of time in the vicinity of the liquidus line, the melt temperature increase causes the relaxation time to decrease. The mechanism of non-monotonic relaxation in non-equilibrium melts which involves the joint influence of non-equilibrium atomic micro-groups dispersion and dissociation on viscosity has been suggested.

**Asknowlegments**

This work was supported partly by the Grant of the Fundamental Research Program of Ural Branch, Russian Academy of Sciences (Project № 12-P-2-1044), and by the by the Russian Foundation for Basic Research (Grant No. 13-02-01149).

**Fig. 1.** DTA thermo-gram of the $Al_{87}Ni_8Y_5$ alloy on heating (*1*) and a typical temperature dependence of the decrement of suspension system viscometer (*2*) when its melting.

**Fig. 2.** $Al_{87}Ni_8Y_5$ viscosity polytherm: curves *1* and *2* denote heating and further cooling with isothermal exposures at each temperature for 30 minutes; *3* denotes heating after a long isothermal exposure (300 minutes) at 870°C; *4* denotes cooling after the isothermal exposure (90 minutes) at 1100°C

**Fig. 3.** The time dependencies of the liquid melt viscosity, $Al_{87}Ni_8Y_5$, at 900°C (*1*), 1050°C (*2*) and 1200°C (*3*), obtained after heating from room temperature

**Fig. 4.** The time dependencies of the liquid melt viscosity, $Al_{86}Ni_8La_6$ (*1*) and $Al_{86}Ni_8Ce_6$ (*2*), at 1100°C

**Fig. 5.** The temperature dependencies of the liquid melt viscosity, $Al_{86}Ni_8La_6$ and $Al_{86}Ni_8Ce_6$, obtained on cooling after long time exposures at 900°C (*1*), 1000°C (*2*), 1100°C (*3*), 1050°C (*4*), and 1200°C (*5*)

**Fig. 6.** Typical temperature dependencies of Al-Ni melts viscosity: *1* is heating, *2* is cooling

**Fig. 7.** The time dependencies of the $Al_{95.5}Ni_{4.5}$ melt viscosity at 760°C (*1*) and 960°C (*2*)

**Fig. 8.** The temperature dependencies of the Al-Y viscosity: *1* is heating, *2* is cooling, *3* are viscosity values, in the samples being isothermally held for a long time

**Fig. 9.** The time dependencies of $Al_{99}Y_1$ viscosity at 700°C (line *1*), $Al_{95}Y_5$ at 870°C (*2*), 1070°C

(*3*) and Al$_{90}$Y$_{10}$ at 1100°C (*4*) and 1200°C (*5*)

**Fig. 10.** The qualitative form of the time dependencies of viscosity of the non-equilibrium melts, resulting in the proposed model for different temperatures, T$_1$ and T$_2$.

Table. The alloys liquidus temperature, relaxation times at various melt temperatures and the parameters approximating the temperature dependences of viscosity of equations that were obtained during cooling after isothermal holding

| Alloys | $t_L$, °C ± 10 °C | t, °C | τ, min | A×$10^{-7}$, $m^2$/sec | $E_v$×$10^3$, J/mol |
|---|---|---|---|---|---|
| $Al_{87}Ni_8Y_5$ | 870 | 900 | 210 ± 20 | 1.29 | 9.23 |
|  |  | 1050 | 130 ± 20 |  |  |
|  |  | 1200 | 75 ± 20 |  |  |
| $Al_{86}Ni_8La_6$* | 860 | 900 | 250 ± 25 | 0.94 | 16.6 |
|  |  | 1000 | 180 ± 20 |  |  |
|  |  | 1100 | 130 ± 10 |  |  |
|  |  | 1200 | 110 ± 10 |  |  |
|  |  | 1300 | 70 ± 10 |  |  |
| $Al_{86}Ni_8Ce_6$* | 885 | 900 | 200 ± 20 | 1.28 | 13.2 |
|  |  | 1000 | 160 ± 10 |  |  |
|  |  | 1100 | 135 ± 10 |  |  |
|  |  | 1200 | 80 ± 10 |  |  |
| $Al_{86}Ni_6Co_2Gd_4Y_2$* | 905 | 920 | 280 ± 30 | 0.88 | 20.7 |
|  |  | 1100 | 130 ± 10 |  |  |
|  |  | 1200 | 85 ± 10 |  |  |
| $Al_{86}Ni_6Co_2Gd_4Tb_2$* | 915 | 920 | 220 ± 20 | 0.98 | 15.7 |
|  |  | 1100 | 125 ± 10 |  |  |
|  |  | 1200 | 80 ± 10 |  |  |
| $Al_{99}Y_1$ | 650 | 700 | 120 ± 20 | 1.7 | 9.1 |
|  |  | 800 | - |  |  |
| $Al_{95}Y_5$ | 740 | 870 | 160 ± 20 | 1.4 | 13.4 |
|  |  | 1070 | 80 ± 20 |  |  |
| $Al_{90}Y_{10}$ | 920 | 1000 | 200 ± 20 | 1.0 | 17.7 |
|  |  | 1200 | 130 ± 20 |  |  |

* The relaxation times of the melts were determined in [9].

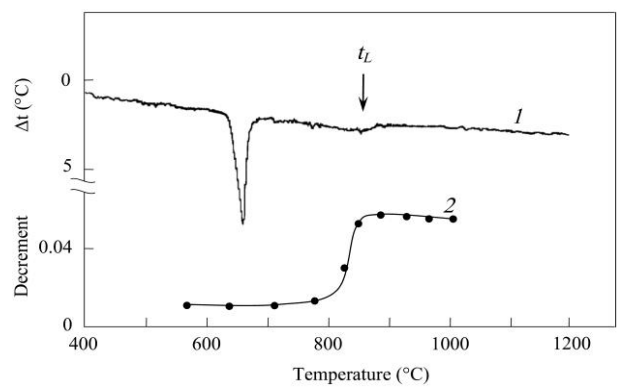

Fig. 1.

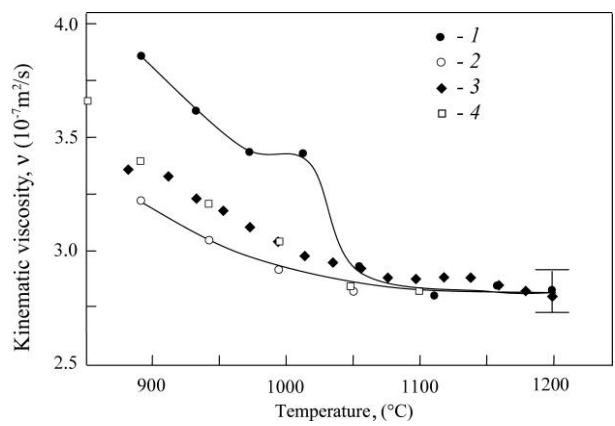

Fig. 2.

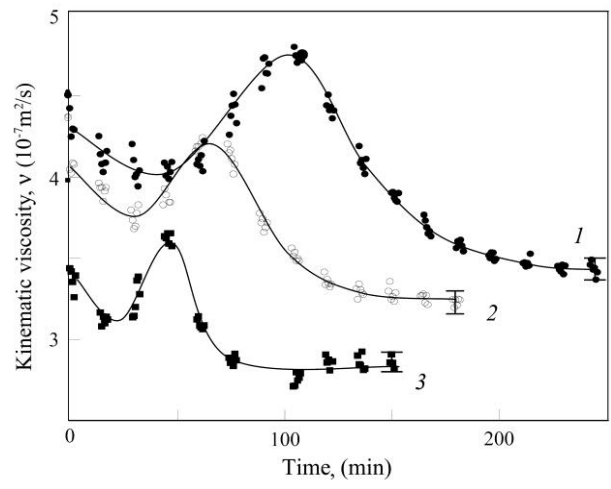
Fig. 3.

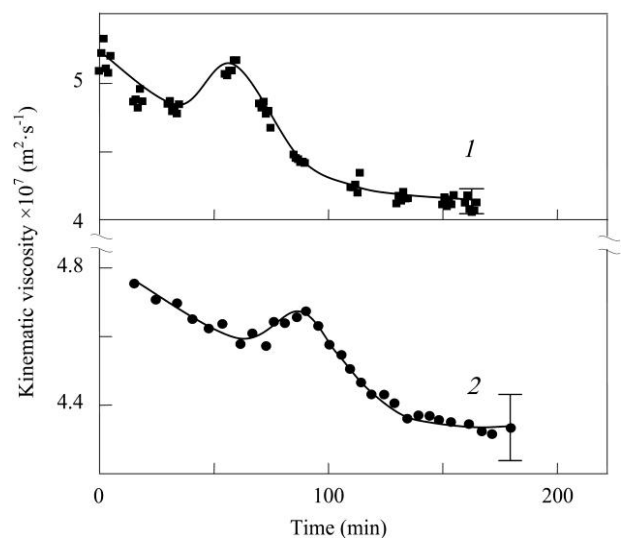

Fig. 4.

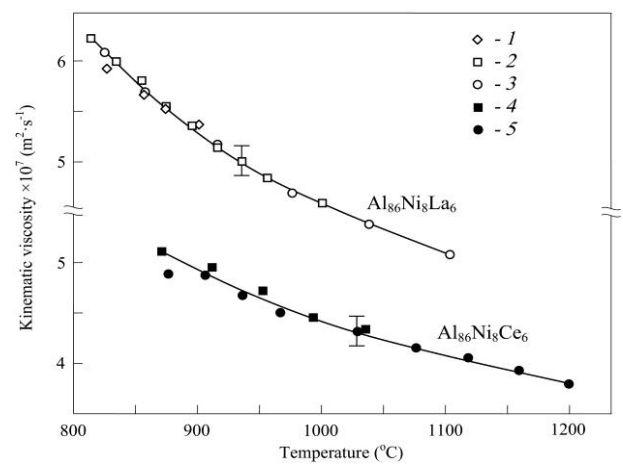

Fig. 5.

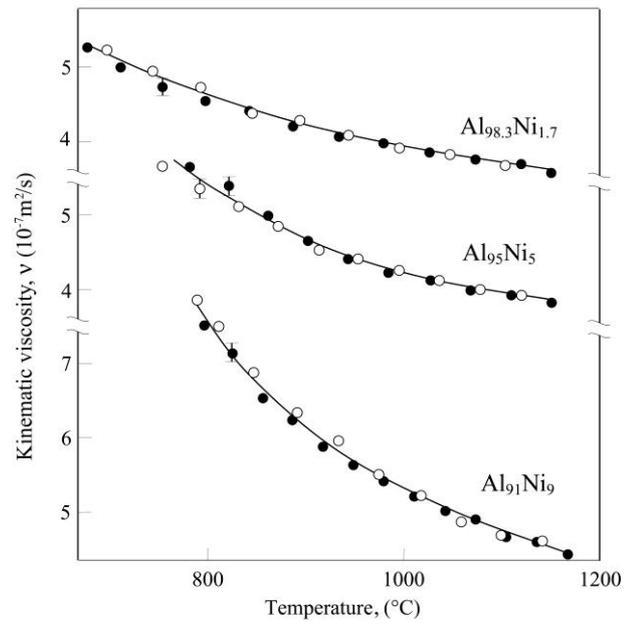

Fig. 6.

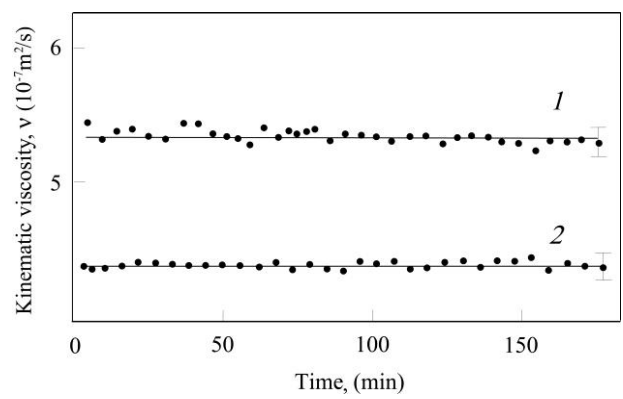

Fig. 7.

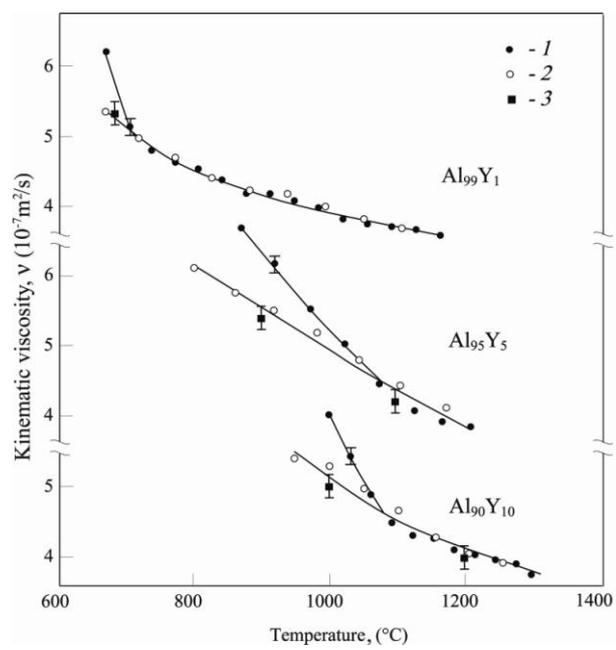

Fig. 8.

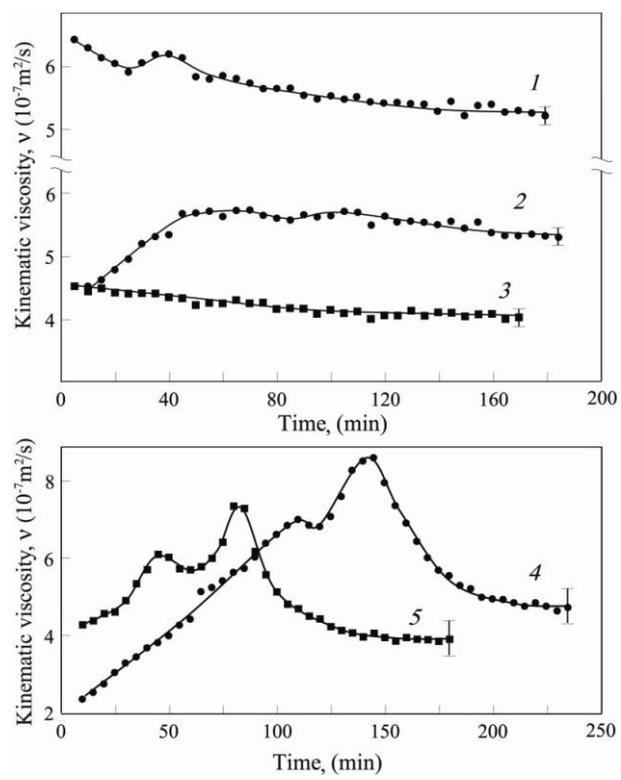

Fig. 9.

**Confirmation of Authorship**

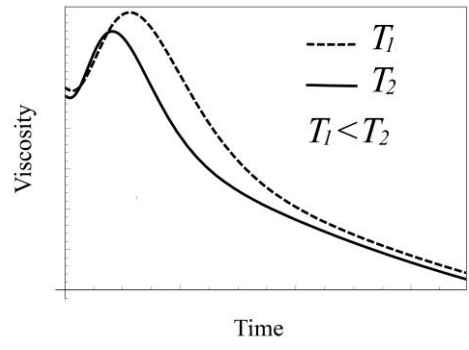

Fig. 10.